\documentstyle[11pt,newpasp,twoside,epsf]{article} 
\def\edcomment#1{\iffalse\marginpar{\raggedright\sl#1\/}\else\relax\fi}
\begin{document} 
\title{Extended Red Emission: Photoluminescence by Interstellar Nanoparticles} 
\author{Adolf N.  Witt} 
\affil{Ritter Astrophysical Research Center, The University of Toledo, Toledo, OH 43606 USA} 
\author{Uma P.  Vijh} 
\affil{Ritter Astrophysical Research Center, The University of Toledo, Toledo, OH 43606, USA}

\begin{abstract}
Interstellar dust in nebulae and in the diffuse interstellar medium of galaxies 
contains a component which responds to illumination by 
ultraviolet photons with
efficient luminescence in the 500 nm to 1000 nm spectral range, known as {\em Extended 
Red Emission} (ERE). We review the techniques of 
detection
and the observational characteristics of the ERE in a wide range of astrophysical 
environments. We then discuss results of the analysis 
of ERE
observations, leading to a set of specific constraints that any proposal for the ERE 
carrier must confront. Finally, we review specific 
models that have been advanced over the past two decades to explain the ERE 
phenomenon. Despite
promising progress on several fronts, no completely satisfactory model for the ERE 
carrier/process exists at this time.
\end{abstract}

\section{Introduction}
Our knowledge of the existence of interstellar dust, its spatial distribution, its 
chemical composition and size distribution, and its 
mass relative to the rest of the interstellar medium is almost exclusively based upon 
observations of the interactions of interstellar 
dust grains with radiation.
The processes studied in the past included mainly interstellar extinction and
polarization of starlight at optical, near-IR and ultraviolet wavelengths, followed by 
studies of dust emission over a wide range of 
wavelengths from the near-IR to
the sub-mm and microwave region. In this paper we are reviewing efforts to study 
interstellar dust through the process of 
photoluminescence (PL), a 
process in which absorptions of photons at ultraviolet/optical wavelengths are 
followed by electronic transitions associated with the 
emission of longer-wavelength optical and near-IR photons.
\begin{figure}[ht]
\plotone{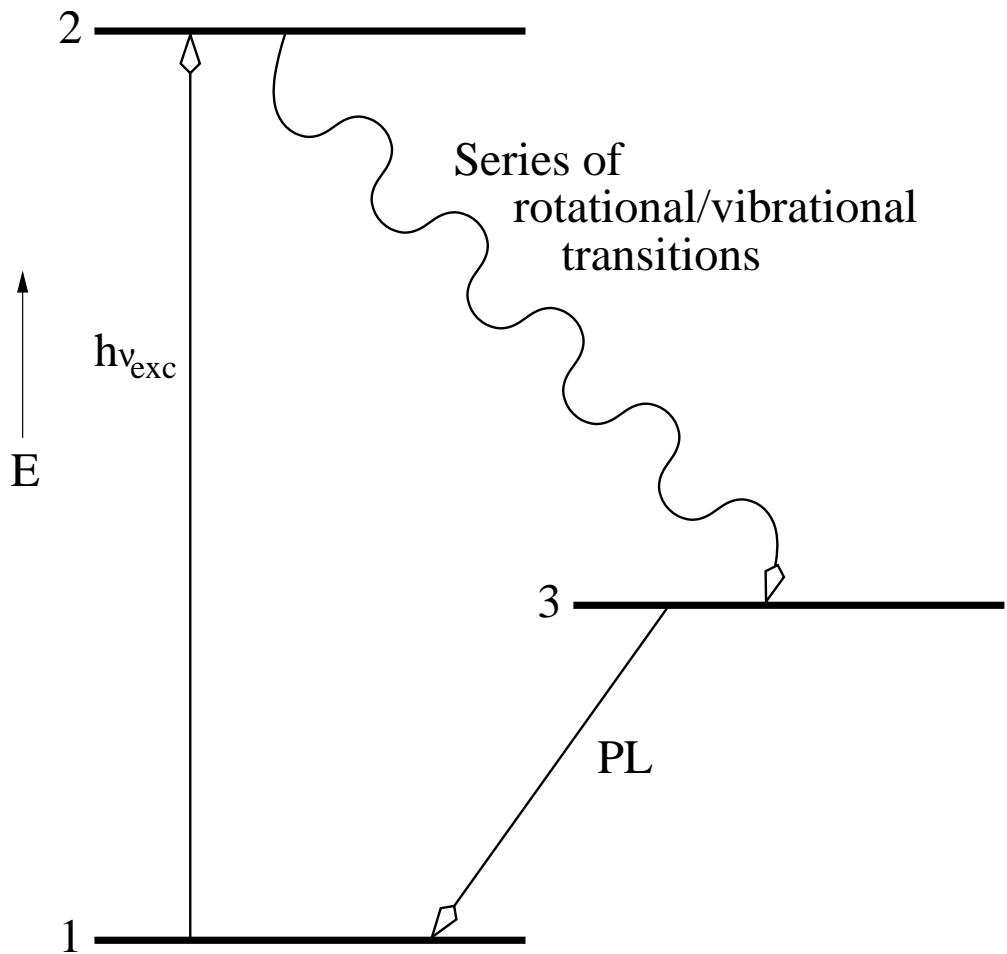}
\caption{Schematic energy level diagram of a photoluminescing system}
\end{figure}
Figure 1 schematically illustrates the physics of a PL process. The ground state (1) 
represents the electronic ground state of a large 
molecule or molecular ion, or a state near the top of the valence band in a 
semiconductor particle. The excitation of the PL process 
under astrophysical conditions results from the absorption of a single UV/optical 
photon, leading to an electronic transition from state 
(1) to state (2). State (2) typically is a bound, high-lying vibrational/rotational 
level of the first or second electronically excited 
state of a molecule or molecular ion, or a high state in the conduction band of a 
semiconductor particle. The excited system relaxes 
through a series vibrational/rotational transitions until the electron finds itself in 
state (3),
from where an optical electronic transition back to the ground state (1) is possible. 
In a polycyclic aromatic hydrocarbon (PAH) molecule, for example, state (3) can either be the lowest 
state in the singlet or triplet vibrational/rotational manifold of the first excited
electronic level. The resulting emissions are referred to as fluorescence and
phosphorescence, respectively. In a semiconductor particle, state (3) is a
state near the lower edge of the conduction band, and the downward transition across 
the band-gap is simply referred to as photoluminescence. For a more complete review of 
PL in aromatic molecules and in semiconductor nanoparticles, the reader should consult the works of Birks (1970) and Yoffe (2001).

Different photoluminescing systems can be characterized by the energy difference 
between the exciting and PL photons, known as 
the Stokes shift, and by the quantum yield or photon conversion efficiency, measured 
by the number ratio of PL photons and exciting 
photons needed to produce the photoluminescence. Most luminescing systems exhibit 
quantum yields of substantially less than 100$\%$, but if the 
energy of the exciting photon is more than twice the energy of the PL photon, highly 
isolated molecules or molecular ions are expected to 
yield two (or more) PL photons per excitation, in a process known as Poincar\'{e} 
fluorescence (Leger et al. 1988). Thus, quantum yields 
well in excess of 100$\%$ may result in systems with large Stokes shifts
under astrophysical conditions, although no specific identifications invoking this 
process have been made. In most natural systems, ionization or the creation of defects   
either quenches or shifts the PL into another spectral region, e.g. the near-IR, which permits 
relaxation to occur without the emission of an optical 
photon, thus reducing the quantum yield to well below $100\%$.

Early suggestions that the high surface brightness of reflection nebulae
in the visible might be due in part to fluorescence by the nebular dust grains
were advanced by Struve \& Swings (1948) and by Aller (1956). Indeed, many minerals 
now thought to be part of the composition of 
interstellar grains do fluoresce under illumination by ultraviolet light or following 
particle bombardment (Blair \& Eddington 1968; 
Koike et al. 2002). However, a first sensitive search for the
existence of such fluorescence in several bright reflection nebulae at wavelengths 
shortward of $490\ $nm by Rush \& Witt (1975) failed 
to reveal any
detectable sign of dust fluorescence, leading to the conclusion that the
nebular surface brightness at these wavelengths was a result of scattering with a 
relatively high grain albedo.

The advent of new, sensitive detectors in the red region of the visible spectrum
in the 1970's brought the first detections of PL, although they were not immediately 
recognized as such.
The detection of the broad luminescence band in the spectrum of
the Red Rectangle nebula (Cohen et al. 1975) was followed by its analysis in terms of 
unusual dust scattering properties by Greenstein \& 
Oke (1977), although the possibility of fluorescence was mentioned. Subsequently, with 
much better
observational data, Schmidt, Cohen, \& Margon (1980) suggested a molecular-emission 
origin for the broad band of excess radiation in the 
Red Rectangle.
The belief that this object was unique in exhibiting this excess radiation
and the lack of a context of other sources showing the same phenomenon
clearly contributed to our inability to recognize the true nature of the
luminescence process in the Red Rectangle. For example, the spectrum of red
excess radiation detected in the Galactic dark nebula Lynds 1780 by Mattila (1979) 
appeared sufficiently different that its connection to 
the Red Rectangle
was not recognized until several years later (Chlewicki \& Laureijs 1987).

A breakthrough occurred in this field with the realization that dust luminescence in 
the red part of the spectrum, extended red emission 
or ERE, as it soon became known, was a common feature in many dusty environments that 
are illuminated by ultraviolet photons. Detections 
of the ERE in reflection nebulae
(Witt, Schild, \& Kraiman 1984; Witt \& Schild 1986, 1988) soon led to the observation 
of ERE in high-latitude cirrus clouds 
(Guhathakurta \& Tyson 1989) and to the discovery of the presence of ERE in the 
continuum spectrum of
planetary nebulae (Furton \& Witt 1990, 1992). Perrin \& Sivan (1992) started a series 
of discoveries of the ERE band in HII regions with 
the Orion nebula,
followed by the 30 Doradus nebula in the LMC (Darbon, Perrin, \& Sivan 1998)
and the HII region Sh 152 (Darbon et al. 2000). The ERE detections in other external 
galaxies include the halo of M82 (Perrin, Darbon, \& 
Sivan 1995) and the
prominent dust lane in NGC 4826, the Evil Eye galaxy (Pierini et al. 2002).
The detection and measurement of ERE in the diffuse interstellar medium
of the Milky Way Galaxy over a wide range of Galactic latitudes by Gordon, Witt, \& 
Friedmann (1998) represents a particular milestone 
for three reasons.
It demonstrated that the ERE carrier is a component of interstellar dust on a 
Galaxy-wide scale; it represented detections of the ERE 
intensities at a level
three orders of magnitude fainter than observed in the Red Rectangle; and finally, it 
permitted the first reliable estimate of the 
quantum yield or
photon conversion efficiency of the ERE process.

In this paper we review first the observational techniques used for the detection, 
independent confirmation, and study of the ERE in 
different types of sources. We then discuss the analysis of the existing ERE data and 
the environments in which ERE is found, leading to 
a set of observational constraints that specific proposals for ERE carriers must 
confront.
Finally, we will introduce specific models that have been advanced over the
past two decades in order to explain the ERE phenomenon.

\section{Observational Techniques for ERE Detection}
Several factors contribute to the challenges one faces when attempting to observe ERE. 
First, ERE appears only in spatially extended 
objects, usually of relatively low surface brightness. Second, ERE is a very broad 
emission, spanning the wavelength range from 540 nm to 
at least 950 nm. And third, it always occurs in conjunction with other sources of 
diffuse emission or scattered light, most of which 
frequently are much brighter than the ERE. For example, in reflection nebulae the 
competing radiation is dust-scattered light; in HII 
regions dust-scattered light, atomic recombination continua as well as line emissions 
occur in the same spectral region, while in 
external galaxies integrated starlight is a contender as well. Attempts to measure ERE 
as part of the high-Galactic-latitude background 
require careful subtractions of atmospheric foregrounds such as airglow and 
atmospheric scattered light, zodiacal light, integrated 
starlight and diffuse Galactic light, before the ERE can be isolated. As a result, 
different observational techniques must be applied to 
different types of sources.
\subsection{Color-Difference Technique}
The color-difference technique is an efficient way to probe the presence of ERE in 
extended objects, provided that  broadband 
colors of the source are readily predictable in the absence of ERE, and the presence 
of sufficient ERE changes these colors. This is the 
case when the competing radiation is dust-scattered starlight, e.g. in reflection 
nebulae (RN) or in the the diffuse Galactic background 
radiation. In an optically thin RN, the well-known wavelength dependence of the 
scattering optical depth leaves the nebular radiation 
bluer than that of the illuminating star throughout the optical spectrum. A simple 
approach for estimating the color difference between 
scattered and illuminating radiation is given by Witt (1985). If, therefore, the V-R 
color of a nebular region is redder than that of the 
star while the B-V color is bluer, this is a strong indication for the presence of 
ERE. This can be further confirmed by spectroscopic 
techniques (see below). Data suitable for the color-difference method can be obtained 
by broadband CCD imaging of individual RN (Witt \& 
Schild 1985, 1986). In Figure 2, we show how the color differences $\Delta$C(V,R) and 
$\Delta$C(V,I) between nebular regions in the RN 
NGC 2023 and NGC 7023 and their respective illuminating stars are systematically too 
red for their corresponding $\Delta$C(B,V) color 
differences.
\begin{figure}
\plotone{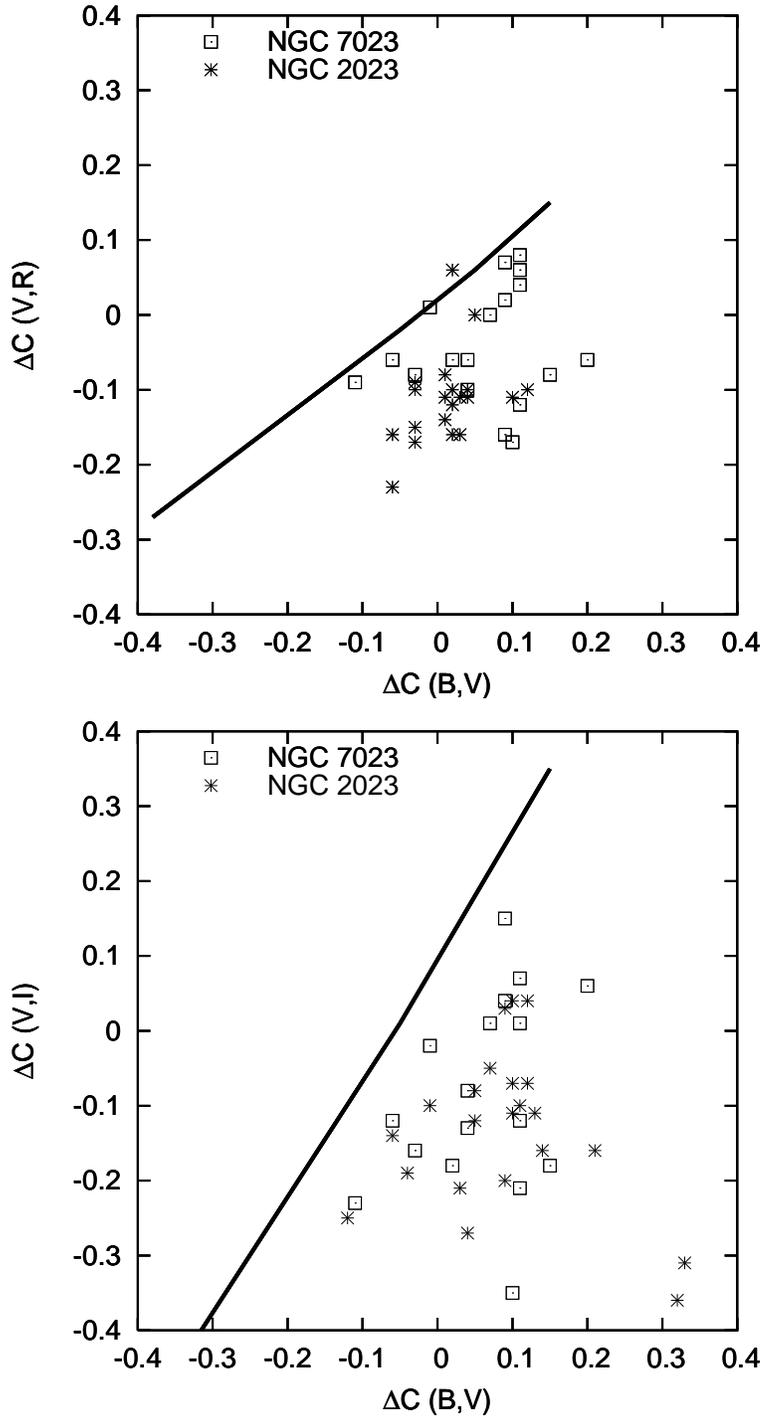}
\caption{The color differences between nebular regions and the illuminating stars are 
shown for the reflection nebulae NGC 2023 and NGC 
7027. Any color difference is defined as $\Delta\textrm{C}(\lambda_1,\lambda_2) = 
\log(S_N/F_*)_{\lambda_1} - \log(S_N/F_*)_{\lambda_2}$. 
The solid lines represent the loci of expected data for scattering alone. The 
observations indicate excess nebular surface brightness 
$S_N$ in the R- and I-bands.}
\end{figure}
The degree to which each measured point falls below the expected relations between the 
color differences  based on scattering alone 
(solid line) is a measure of the relative strength of the ERE. These observational 
data show that ERE is present to variable degrees in 
both the R- and I-bands in different locations within NGC 2023 and NGC 7023. Early
detections of ERE in high-latitude cirrus clouds 
(Guhathakurta \& Tyson 1989; Gordon et al. 1998) also relied upon the fact that the 
color of the cirrus is substantially redder than that 
of the illuminating starlight. 
\subsection{Long-Slit Spectroscopy}
The spectroscopy of ERE in low-surface-brightness, extended sources is strongly
affected by terrestrial airglow. Long-slit spectrographs provide the best approach to 
simultaneously record and subsequently subtract 
the foreground sky spectrum.
Also, in sources whose spectrum is dominated by emission lines, e.g. HII regions and 
planetary nebulae, long-slit spectroscopy offers a 
possibility for studying
the faint continuum, including the broad ERE band, between strong emission lines. The 
large extent of the ERE in wavelengths requires a 
combination of relatively low wavelength resolution and broad spectral coverage. If 
the resulting spectrum is a combination of scattered 
light and ERE, as for example in RN,the presence and relative strength of the ERE is 
revealed by dividing the nebular spectrum by the 
spectrum of the illuminating star, as shown in Figure 3. 
\begin{figure}
\plotone{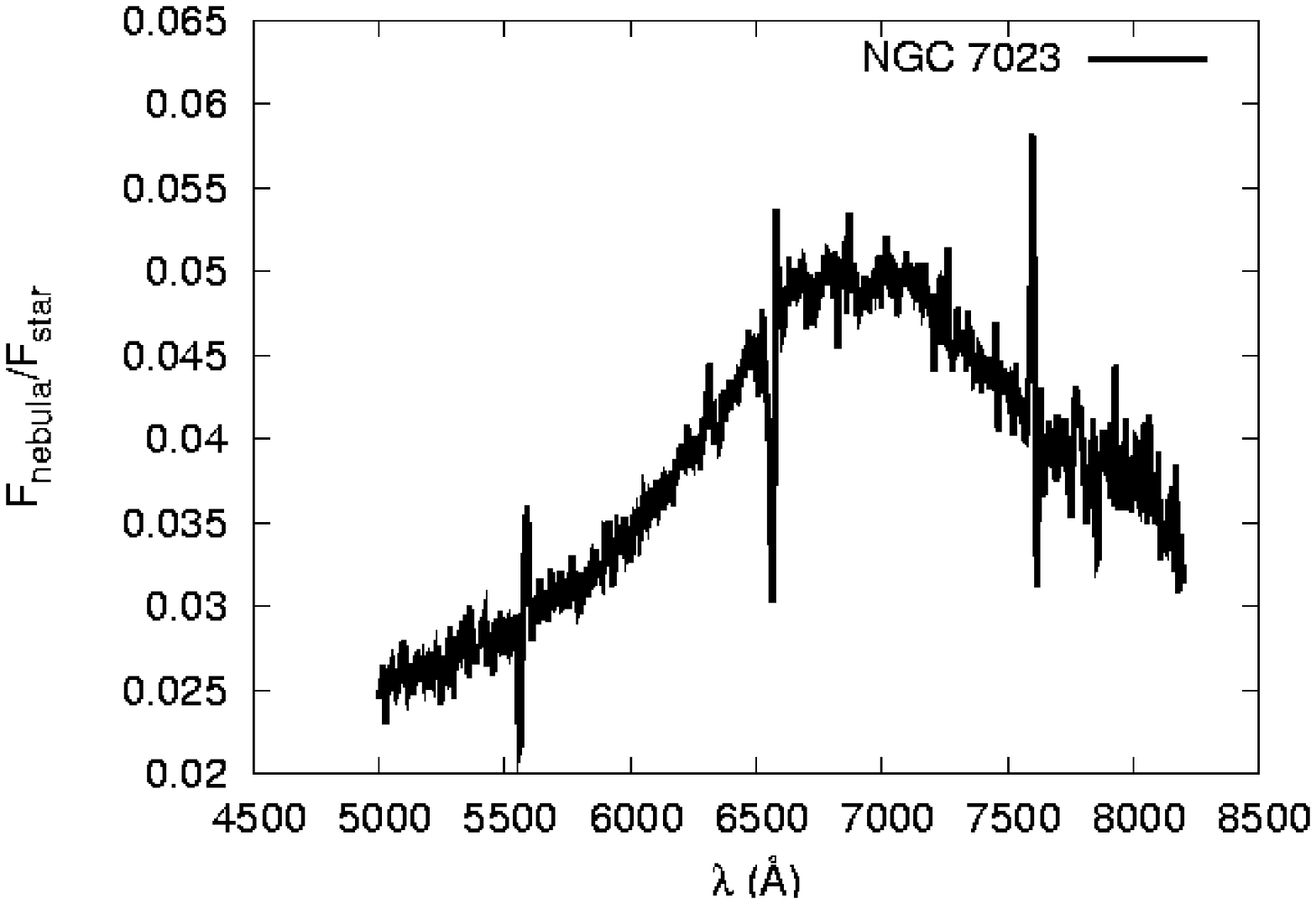}
\caption{Ratio of nebular to stellar spectrum in the SW filament of NGC 7023. The 
sharp features near 5600 \AA, 6560 \AA, and 7600 \AA\ 
are a result of incomplete cancellation of night sky and nebular emission features and 
are not characteristics of the ERE.}
\end{figure}
The application of long-slit spectroscopy to the search for ERE in planetary nebulae 
is described by Furton \& Witt (1990; 1992), and the 
use of the same method in HII regions was first demonstrated by Perrin \& Sivan (1992) 
in the case of the Orion nebula. ERE spectra and 
band-integrated intensities as well as observed upper limits for a large sample of 
reflection nebulae based on 
long-slit spectroscopy have been reported by Witt \& Boroson (1990). 
\subsection{Line-Depth Technique}
The presence of quasi-continuous luminescence may be difficult to discern, if
it lacks a distinctive spectral feature, as is the case with the ERE band. In 
reflection nebulae, however, one can easily distinguish 
dust luminescence from scattered radiation by virtue of the fact that scattering 
faithfully reproduces relative line depths or equivalent 
widths of spectral lines present in the spectrum of the illuminating source. The 
presence of a certain fraction of luminescence at the 
positions of such lines is revealed by a proportional reduction in the relative line 
depths in the nebular spectrum compared to 
corresponding relative line depths in the illuminating star, as illustrated in Figure 
4.
\begin{figure}
\plottwo{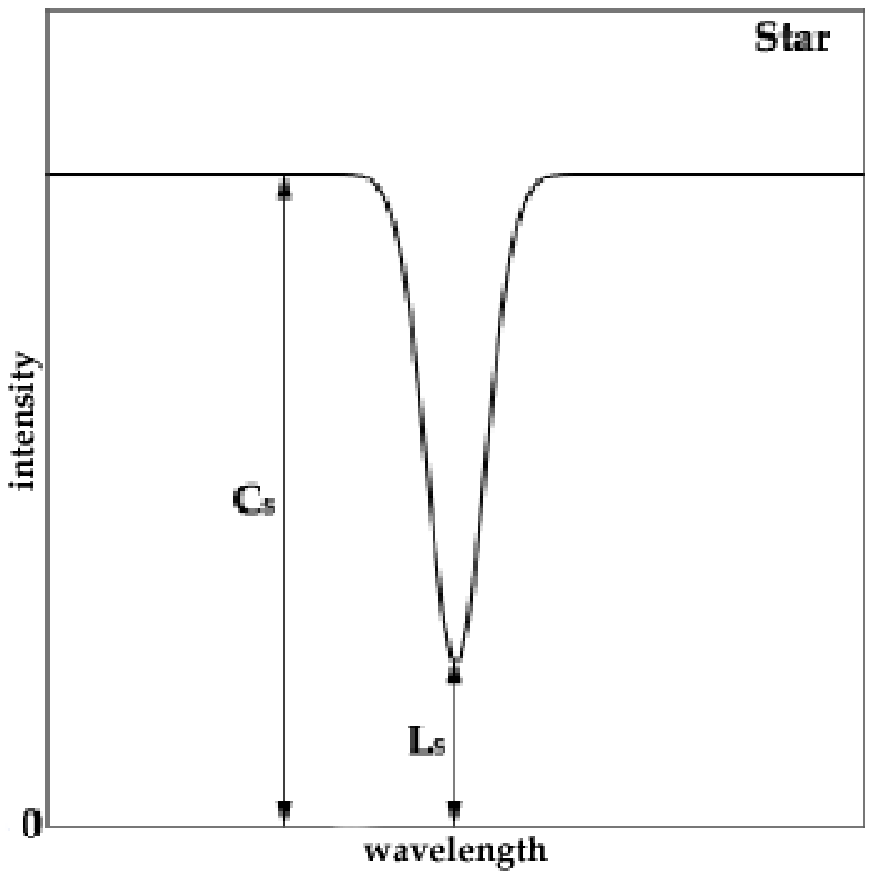}{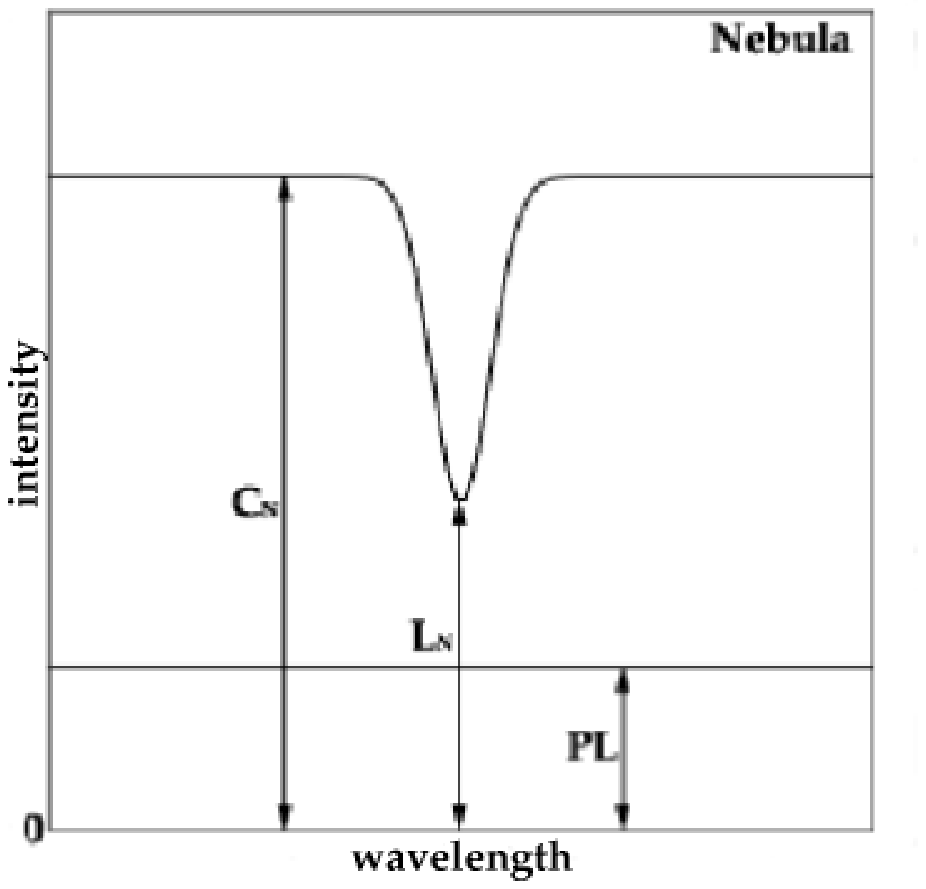}
\caption{Illustration of the line-depth technique. In the nebular spectrum, consisting 
of scattered light and PL, the relative line depth 
$(C_N-L_N)/C_N$ is smaller than the corresponding value $(C_S-L_S)/C_S$ in the stellar spectrum.}
\end{figure}
This method is particularly useful in the search for luminescence in the blue part of 
the spectrum, where strong hydrogen Balmer lines 
dominate the spectra of most reflection nebulae. It is important to note that the spectral 
resolution of both the stellar and the nebular spectrum must be identical to avoid a false-positive
result. This method was first employed by 
Rush \& Witt (1975), who used photographic 
spectroscopy and narrow-band photoelectric photometry to place useful upper limits on 
the contribution made by possible dust 
photoluminescence to the observed surface brightness of several reflection nebulae. A 
recent application of this technique by Vijh, Witt, 
\& Gordon (2003) to sources with known strong ERE, now based on long-slit CCD 
spectroscopy, resulted in much tighter upper limits 
consistent with zero luminescence at wavelengths of H$\beta$ and shorter.
\subsection{Unsharp Masking}
The data shown in Figure 2 demonstrate the high degree of spatial variability
of the ratio of ERE to scattered light within a single nebula. A closer examination of 
the ERE distribution in NGC 2023 and NGC 7023 
reveals that the ERE intensity is strongly enhanced in narrow filamentary regions 
situated on the surfaces of dense molecular clumps 
facing the respective illumination source.
These bright structures are a result of a favorable viewing geometry and a high
local opacity for the stellar radiation responsible for ERE excitation, thus leading 
to exceptionally large column densities of ERE 
emitters, confined to a space with narrow lateral dimensions. The resulting ERE 
structures can be imaged
with the unsharp masking technique (Malin \& Zealy 1979), which permits the 
simultaneous display of sharp features ranging in intensity over four orders of 
magnitude. Witt \& Malin (1989) applied this technique 
with good success to
NGC 2023. The location, width and orientation of numerous ERE filaments was revealed, 
enabling subsequent studies of spatial correlations 
with filamentary emissions by fluorescent molecular hydrogen (Field et al. 1998) and 
by CN (Fuente, Martin-Pintado, \& Gaume 1995).
\subsection{Imaging Spectropolarimetry}
Scattered light in reflection nebulae is substantially linearly polarized at optical 
and near-IR wavelengths, typically at the 20\% to 
30\% level. Watkin, Gledhill, \& Scarrott (1991), through the use of imaging 
polarimetry,  discovered that localized ERE filaments in NGC 
7023 are associated with a significant reduction of linear polarization in the R- and 
I-bands, consistent with a dilution of polarization 
through the presence of unpolarized emission with an intensity equal to that of the 
ERE, as determined by independent spectroscopic and 
photometric techniques. This demonstration that ERE is unpolarized makes attempts to 
explain the ERE in terms of scattering with unusual 
dust characteristics (e.g. Greenstein \& Oke 1977) unsustainable.
Rather, the ERE must be the result of an unpolarized emission process.
\section{Analysis of Observational Data}
The identification of the ERE carrier remains an outstanding challenge. None of the 
current "unified" dust models (e.g. Draine 2004, Dwek 
et al. 2004, this volume), which attempt to satisfy several observational constraints 
simultaneously, predict the existence of the ERE, 
nor did they explain the phenomenon after its discovery. At this stage, an extensive 
set of observational data exist which any ERE 
carrier candidate must meet. Finding some material which produces approximately the 
correct spectrum of some ERE source is simply not 
sufficient. To this end, we now summarize the observational information about the ERE 
and try to deduce the resulting constraints for 
models.
\subsection{ERE Environments}
The ERE has been observed by one or more of the techniques described above in the 
following astrophysical environments:
\begin{enumerate}
\item reflection nebulae
\item HII regions
\item dark nebulae (illuminated by the ISRF)
\item high-latitude Galactic diffuse ISM and cirrus
\item dusty ISM of external galaxies
\item carbon-rich planetary nebulae
\item the Red Rectangle, a peculiar proto-planetary nebula
\end{enumerate}
These environments have two factors in common: they contain dust and they are
illuminated by radiation fields containing UV photons. In terms of UV photon 
densities, these regions span a range of six orders of 
magnitude. In terms of
dust, both well-mixed and processed interstellar dust mixtures (the first five cases) 
and regions where dust is produced in local 
outflows (the last two cases) are represented.
The data suggest that once formed, the ERE carrier is fairly robust and can survive 
under a wide range of astrophysical conditions. The 
fact that evidence for ERE was found in carbon-rich planetary nebulae and not in 
oxygen-rich nebulae, despite comparable efforts on both 
types of objects (Furton \& Witt 1992), suggests a carbonaceous nature for the carrier 
of the ERE. This argument
has lost some of its power in light of the {\em Infrared Space Observatory} (ISO) 
observations of carbon-rich planetaries (Waters et al. 
1998), which show the presence of strong spectral features of crystalline silicate 
features in the mid-IR spectral region. On the other 
hand, the non-detection of ERE in oxygen-rich planetaries does argue against an 
association of ERE with oxygen-rich materials. This 
conclusion is further supported by the fact that the most prominent ERE source with 
local dust production, the Red Rectangle nebula, is a 
carbon-rich environment. When considering ERE environments, it must be emphasized that 
the presence of dust and UV photons appears to be 
a necessary but not a sufficient condition. A number of bright reflection nebulae 
associated with B-stars as illuminating sources have not 
revealed any evidence of ERE presence (Witt \& Boroson 1990), prominently among them 
the Merope nebula in the Pleiades, often presented 
as the prototype of reflection nebulae.  Further evidence that scattering grains, i.e. 
the large-size end of the grain size distribution, 
are not directly involved with the production of ERE comes from the fact that the ERE 
morphology of the Red Rectangle nebula is 
distinctly different from its scattered-light morphology, and that in planetary 
nebulae, where ERE is detected, no trace of scattered 
light can be observed.
\subsection{Excitation of ERE}
The ERE is caused by PL. This is confirmed by the observation that the {\em maximum} 
ERE intensity in any given environment is closely 
correlated with the density of the local radiation field
(Gordon et al. 1998), in particular the density of UV photons (Smith \& Witt 2002). In 
dense, clumpy nebulae, it is always the side of 
clumps facing the closest source of UV photon that exhibit the highest ERE intensity 
(Witt \& Malin 1989). These facts leave little doubt 
that ERE is powered by the absorption of photons. Several observational studies 
suggest strongly that these photons originate in the 
ultraviolet spectral region.

Darbon, Perrin, \& Sivan (1999), summarizing all successes and failures of searches 
for ERE in reflection nebulae and HII regions, found 
that only environments illuminated by sources as hot as $10^4$K and hotter exhibited 
ERE, while sources with illuminating stars with 
$T_{eff} \leq$ 7000 K did not, although ample dust was present in the nebulae with 
these cooler stars. The results of Darbon et al. 
(1999) are reproduced in Figure 5. 
\begin{figure}
\plottwo{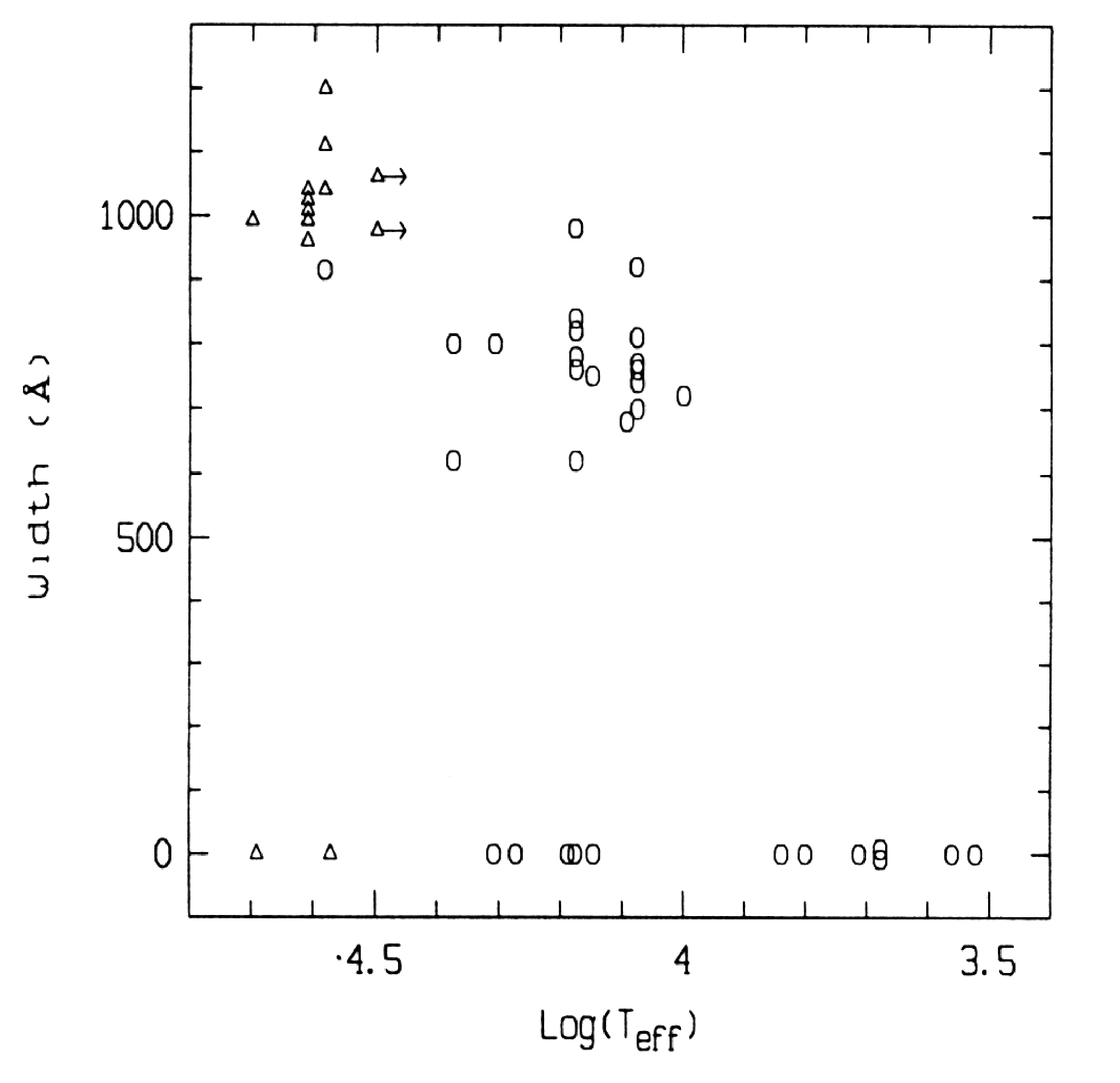}{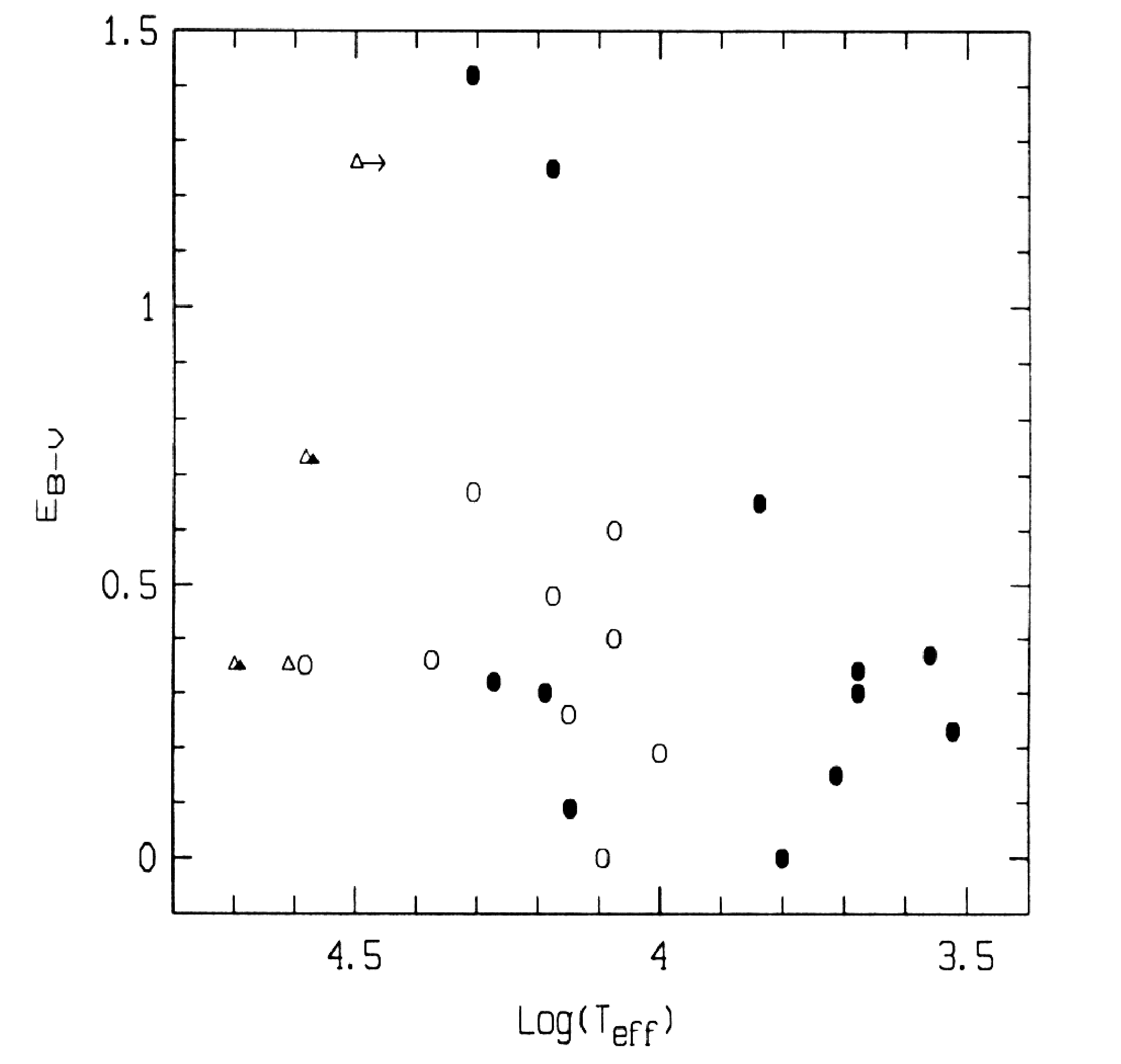}
\caption{Left: Width of ERE bands detected in HII regions ($\triangle$) and in 
reflection nebulae ($\circ$) as function of the effective 
temperature of the exciting stars. Right: Color excess E(B-V) of the illuminating 
stars in HII regions ($\triangle$) and reflection 
nebulae ($\circ$) with positive ERE detections. Filled symbols correspond to sources 
where ERE has not been detected. These figures have 
been reproduced from Darbon et al. (1999), with permission.}
\end{figure}
The results suggest that photons with energies E $\geq$ 7.25 eV, the approximate 
UV-cutoff of $T_{eff}$ = 7000 K atmospheres, are 
required for ERE excitation. This conclusion is supported by observations of HD 44179, 
the exciting star of the Red Rectangle,
with the \emph{International Ultraviolet Explorer} (IUE) satellite by Sitko, Savage, 
\& Meade (1981). The spectrum of HD 44179 displays a 
sharp absorption edge, beginning at 170 nm (7.25 eV) and extending to shorter 
wavelengths, totally inconsistent with normal dust 
extinction. In the light of the Darbon et al. (1999) result and given that the Red 
Rectangle is an exceptionally bright ERE source, it 
seems likely that this far-UV absorption at wavelengths $\leq$ 170 nm is caused by the 
ERE carrier particles. This result is further 
supported by Witt \& Schild (1985), who examined the propagation of ERE-exciting 
photons through the well-studied reflection nebulae NGC 
2023 and NGC 7023.
They found the opacity for the exciting radiation in these two nebulae
to be $A_{exc}$/E(B-V) = 9.16 and 7.15, respectively, with an uncertainty of $\pm$1.5, 
consistent with the dominant absorption of the 
ERE-exciting radiation occurring at wavelengths $\leq$ 150 nm. The determination of 
the spectral region in which the excitation of the 
ERE takes place is a crucial constraint for the identification of the ERE carrier.
\subsection{Variability of the ERE Spectrum}
In general, interstellar absorption and emission features are characterized by the 
relative constancy of their central or peak 
wavelengths. By contrast, the large variability of the peak wavelength of the ERE band 
is a defining characteristic of the ERE process. 
The ERE peak wavelength varies from source to source and within a given source with 
distance from the illuminating star. In the majority 
of sources the ERE peak occurs in the 650 nm to 750 nm range, but extreme cases 
ranging from 610 nm to 950 nm have been found. A positive 
correlation between the width and the peak wavelength of the ERE band has been 
demonstrated by Darbon et al. (1999), with the FWHM of the 
band varying from 60 nm to 120 nm with corresponding changes in the peak position 
between 650 nm and 780 nm. Smith and Witt (2002) have 
shown that the ERE peak wavelength is well correlated with the local density of the 
far-ultraviolet radiation field, as shown in Figure 
6. 
\begin{figure}
\plotone{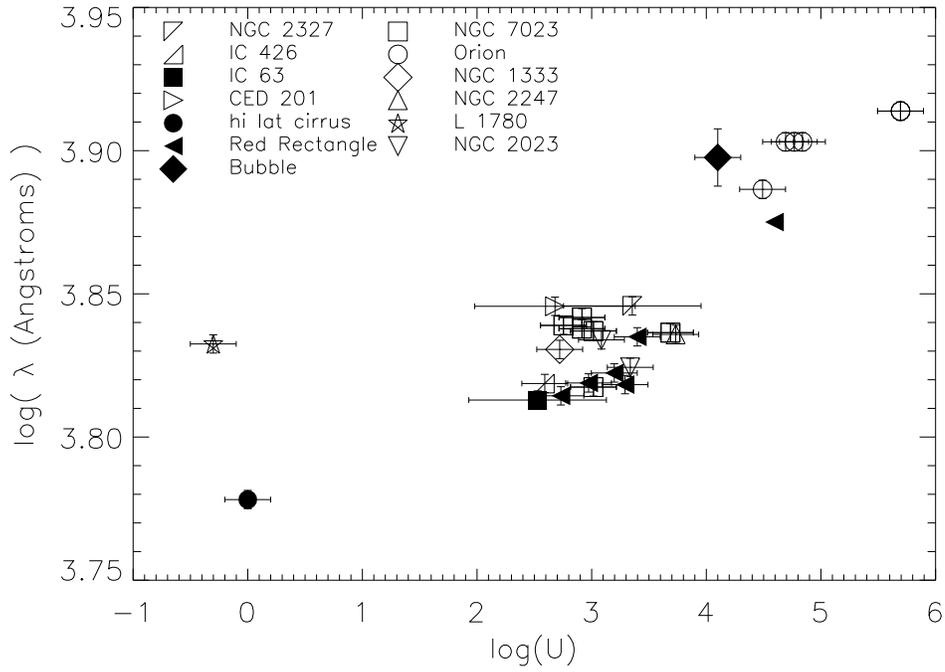}
\caption{The logarithm of the peak wavelength is plotted against the logarithm of the 
density of the far-ultraviolet radiation field (in 
units of the FUV-radiation field density in the diffuse ISM near the Sun) at the 
location of the ERE measurement. Adapted from Smith \& 
Witt (2002).}
\end{figure}
They proposed that a photo-thermo-dissociation process, increasingly effective in 
high-radiation-density environments, selectively 
destroys the smaller ERE carrier particles, shifting the dominant size of the 
particles towards larger sizes. In this model, the emission 
wavelengths are related to the sizes of carrier particles through quantum confinement 
as found in luminescing nanoparticles (Ledoux et 
al. 2002) or large organic molecules (Platt 1956). It is clear from the observational 
evidence that models for the ERE carrier which do 
not account for the variability of the ERE spectrum are not sustainable.
\subsection{Correlations/Anti-correlations with other Interstellar Features}
The quest for the discovery of the ERE carrier could be narrowed substantially, if 
correlations with other interstellar features could be 
established, for which reasonable explanations have already been found. Unfortunately, 
no clear correlations have emerged so far.

Early ERE studies (Witt \& Schild 1988; Witt \& Malin 1989) suggested an association 
between ERE and H$_2$ photo-dissociation fronts, 
which appeared to support models in which processing of amorphous carbon with atomic 
hydrogen led to PL by hydrogenated amorphous carbon 
dust (e.g. Duley \& Williams 1990). Later studies of high-spatial-resolution 
H$_2$-fluorescence maps of NGC 2023 by Field et al. (1994, 
1998) revealed significant anti-correlations with ERE, with ERE being observed in the 
absence of H$_2$-luminescence. Where there is a 
coincidence of ERE filaments and H$_2$ filaments, it probably results from the fact 
that both luminescence mechanisms require similar 
far-ultraviolet
radiation for their excitation and this radiation penetrates only a small physical 
distance into dense molecular clumps, thus producing 
an apparent strong spatial correlation between two physically unrelated processes.

An equally ubiquitous interstellar component, which is observed in virtually all the 
same environments in which ERE is seen, produces the 
near-IR aromatic hydrocarbon features in emission, also known as the unidentified 
infrared (UIR) bands (see articles by E. Peeters and T. 
Onaka in this volume). d'Hendecourt et al. (1986) and Leger, Boissel, \& d'Hendecourt 
(1988) proposed that fluorescence
by isolated polycyclic aromatic hydrocarbon (PAH) molecules and their ions could 
indeed be the source of the ERE. Following excitation by 
ultraviolet photons, these systems are expected to relax through a combination of 
electronic and vibrational transitions, which would 
suggest correlations between UIR band emission and the ERE. However, spatially 
resolved spectroscopy of ERE in the Red Rectangle (Schmidt 
\& Witt 1991) and of the 3.3 $\mu$m UIR band along an identical slit position (Kerr et 
al. 1999) produced a strong anti-correlation.
If the same carrier is responsible for both emission processes, then this observation 
suggests that electronic and vibrational 
de-excitations must occur under mutually exclusive sets of physical conditions. This 
conclusion was further supported by recent 
observations of ERE and the UIR bands at 3.3 $\mu$m and 6.2 $\mu$m in the compact HII 
region Sh 152 by Darbon et al. (2000), who found 
the ERE to be associated with the ionized hydrogen emission while the UIR bands were 
seen predominantly in the outer, molecular regions. 
This agrees with earlier results by Perrin \& Sivan (1992),
who found no correlations between ERE and UIR bands in the Orion nebula.

Finally, several recent models for interstellar dust attribute the 2175 \AA\
ultraviolet extinction feature either in part or totally to absorption by large PAH 
molecules or clusters. Some specifically predict that 
this absorption provides the energy source for the ERE (Seahra \& Duley 1999; Duley \& 
Seahra 1999) as well as the UIR band emission. It 
is interesting in this context that the nebula with the most intense ERE, the Red 
Rectangle, produces only a very weak 2175 \AA\ feature 
(Sitko et al. 1981), while another nebula, IC 435,
illuminated by a hotter star, exhibiting one of the strongest 2175 \AA\ features
observed (Witt, Bohlin, \& Stecher 1984), shows no observable ERE (Witt \& Boroson 
1990). There is no observational support for the 
proposed connection between the 2175 \AA\ band and ERE.
\subsection{ERE Photon Conversion Efficiency}
The detection of ERE in the high-latitude diffuse Galactic ISM (Guhathakurta \& Tyson 
1989; Gordon et al. 1998; Szomoru \& Guhathakurta 
1998) represents a major breakthrough: It allows the measurement of the ERE intensity 
under optically thin conditions for both the 
exciting and luminescence radiation in an environment where the total dust column 
density can be estimated, e.g. from HI column densities 
or from galaxy counts. Given a fair knowledge of the Galactic interstellar radiation 
field and its spectrum (Mathis, Mezger, \& Panagia 
1983),
it then becomes possible to estimate a lower limit to the ERE quantum yield,
assuming that all absorption at ultraviolet and optical wavelengths by dust along a 
given line of sight is due to the ERE carrier 
particles. Based on observations ranging over a wide range of Galactic latitudes, 
Gordon et al. (1998) found such a lower limit to the 
photon conversion efficiency of the
process leading to ERE of (10$\pm$3)$\%$, a value confirmed by Szomoru \& Guhathakurta 
(1998). Normalized to the HI column density, this 
corresponds to an ERE intensity of ($1.43\pm0.31) \times 10^{-29} \textrm{ ergs 
s}^{-1} \textrm{\AA}^{-1} \textrm{sr}^{-1} 
\textrm{H-atom}^{-1}$. As discussed in Section 3.2 above, only photons at wavelengths 
shortward of 170 nm are likely causes of ERE; in 
addition, there are other known absorbing interstellar dust components not likely 
associated with ERE. For these reasons, this lower 
limit on the ERE quantum yield is extremely conservative, and it is likely that the 
true intrinsic quantum yield of the ERE carriers is 
in the vicinity of 50$\%$ or even higher. Such high quantum yields are commonly found 
in fluorescence of organic molecules (Berlman 1965) 
and in PL by semiconductor nanoparticles (Ledoux et al. 2001). Even with quantum 
yields of the order of $50\%$, the ERE carrier particles would need to absorb about $20\%$ 
of the UV/optical photons. The intensity of the ERE 
in the diffuse ISM requires that the ERE carrier is a dominant dust component causing 
far-UV interstellar absorption, because only a 
relatively small fraction of the absorbed photon energy re-emerges in the form of ERE 
photons. The ERE carriers must therefore consist of 
cosmically abundant refractory elements.

Smith \& Witt (2002) extended the determination of ERE conversion efficiencies
to numerous other sources, where the ERE has been observed superimposed on a 
scattered-light continuum. The latter, in conjunction with 
the spectrum of the illuminating star, was used to obtain an estimate of the local 
radiation density responsible for ERE excitation. The 
Smith \& Witt results are reproduced in Figure 7, where the ERE quantum yield is 
plotted as a function of the local ultraviolet radiation 
density, expressed in units of the interstellar value.
\begin{figure}
\plotone{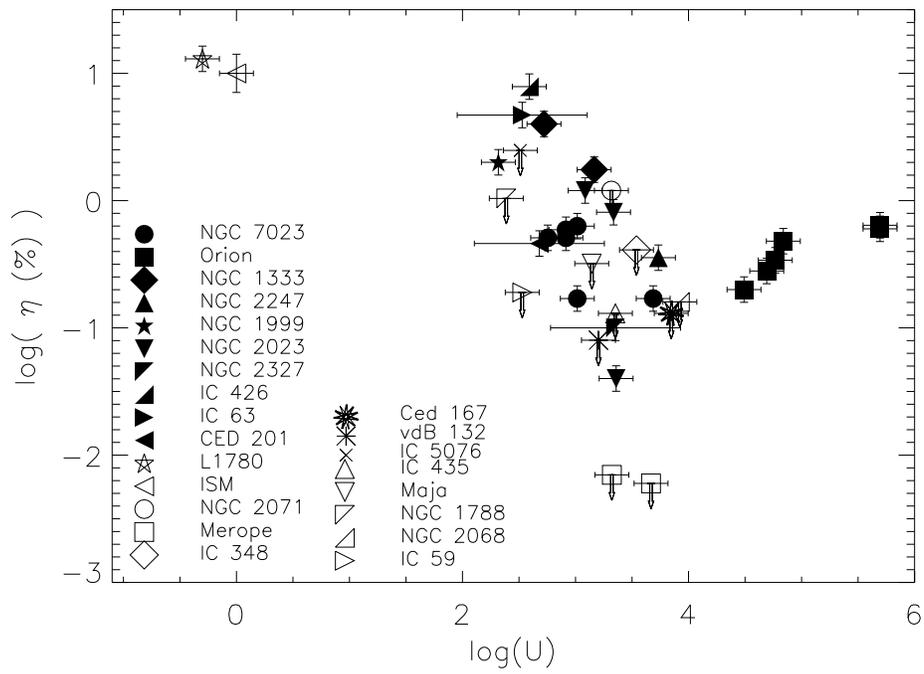}
\caption{The logarithm of the lower limit to the ERE quantum yield $\eta$ is plotted 
against the logarithm of the density of the 
far-ultraviolet radiation field U, measured in units of the far-ultraviolet radiation 
density in the diffuse ISM near the sun. Adapted 
from Smith \& Witt (2002).}
\end{figure}
All quantum yields are lower limits in the sense defined by Gordon et al. (1998).
Consistent with the finding that the ratio of ERE-to-scattered light is highly 
variable within each nebula and from one nebula to 
another, the resulting efficiencies in RN vary over a range of at least two orders of 
magnitude. It is remarkable, however, that the ERE 
efficiency is lower in {\em all} sources when compared to that in the diffuse ISM and 
the high-latitude dark nebula L1780, both 
environments exposed to the low-density interstellar radiation field. Among reflection 
nebulae, there is a distinct downward trend to 
lower ERE efficiencies with increasing radiation field densities. The highest ERE 
quantum yields are found in environments with the 
lowest radiation densities, i.e. the diffuse ISM and surfaces of interstellar dark 
nebulae.
\subsection{Photo-processing of the ERE Carrier}
While the presence of UV photons is essential for stimulating the ERE process in the 
first place, the observations provide strong 
evidence that the interaction of the ERE carrier with radiation brings about changes 
in the carrier with
observable consequences for the ERE spectrum and its relative intensity.
We identify three phenomena which we attribute to photo-processing of the ERE carrier.
\begin{itemize}
\item The peak wavelength of the ERE band shifts to longer wavelengths with increasing 
density of the local radiation field. (Figure 6)
\item In nebulae where ERE is observed, the photon conversion efficiency decreases 
strongly with increasing density of the UV radiation 
field. (Figure 7)
\item Nebulae in which ERE is not detectable (Witt \& Boroson 1990, Table 3)
tend to be diffuse, unclumped environments, in which dust has been fully exposed to 
ultraviolet radiation for a considerable time. In 
clumpy nebulae, such as NGC 2023 and NGC 7023, the ERE efficiency is highest on the 
faces of high-density clumps while decreasing to 
below detectability in low-density interclump regions.
\end{itemize}

This evidence suggests that exposure to increasingly intense radiation fields first 
modifies and then either de-activates or destroys the 
carrier of the ERE.
The details of these processes will depend on the specific nature of the actual 
carrier particles, but any successful model must be able 
to account for this radiation-dependent behavior. Smith \& Witt (2002) noted that 
photoionization
will quench the PL of semiconductor nanoparticles such as
oxygen-passivated silicon nanoparticles (SNPs). The balance between photoionization of 
the SNPs and their recombination with free 
electrons in any given environment would then determine the ERE efficiency. 
Furthermore, in increasingly dense and hard UV radiation 
fields the SNPs would be increasingly multiply ionized, making them subject to 
photo-fragmentation, a process affecting mainly the 
smallest among the carrier particles. The resulting change in the SNP size 
distribution would then lead to the observed shift of the ERE 
peak towards longer wavelengths.
\section{Models for the ERE Carrier}
None of the currently discussed "unified" models for interstellar dust (Draine 2004; 
Dwek et al. 2004, this volume) specifically predict 
or account for the existence of ERE and its observed characteristics. Only one attempt 
(Zubko, Smith, \& Witt 1999) has been made to 
simultaneously meet requirements posed by the wavelength dependence
of extinction, abundance constraints, and ERE observations by including silicon 
nanoparticles as a small-grain component into standard 
dust models. Lack of adequate knowledge of the size-dependent dielectric functions of 
likely materials in the nanoparticle regime is 
currently preventing further work in this direction. Most of the past effort has been 
focused purely on identifying
classes of carrier particles capable of producing the ERE. These have included 
carbon-based solids, such as hydrogenated amorphous carbon 
(HAC) (Duley 1985), quenched carbonaceous composite (QCC) (Sakata et al. 1992), coal 
(Papoular et al. 1996), and bacterial pigments 
(Hoyle \& Wickramasinghe 1999), as well as large carbon-based molecules, such as PAHs 
(d'Hendecourt et al. 1986; Leger et al. 1988), 
$\textrm{C}_{60}$ (Webster 1993) and PAH-clusters (Seahra \& Duley 1999).
Lately, non-carbon bearing ERE candidates such as silicon nanoparticles (Witt et al. 
1998; Ledoux et al. 1998) and silicates (Koike et 
al. 2002) have also received close attention. In order to judge the success or failure 
of these efforts, it is essential to summarize the 
observational constraints such proposals must meet.
\subsection{Requirements}
Here we summarize the conditions imposed by observations on the carrier particles of 
the ERE.
\begin{itemize}
\item ERE is a PL process.
\item Carriers must consist of cosmically abundant refractory elements that are 
depleted from the gas phase in interstellar space and are 
capable of forming photoluminescent materials. Such elements include C, O, Si, Fe, and 
Mg.
\item Carrier particles must survive under a wide range of interstellar and 
circumstellar conditions and have a Galaxy-wide distribution.
\item Efficient excitation of the ERE process requires photons with energies in excess 
of 7.25 eV.
\item No PL is detectable in the 350 to 500 nm wavelength range under excitation by UV 
continuum radiation.
\item The PL occurs in a broad, unstructured band with a peak wavelength varying 
between 600 nm and $>$ 900 nm, within given sources and 
from source to source, in response to increasingly dense and hard illumination by UV 
photons.
\item The width and the peak wavelength of the ERE band are positively correlated.
\item Under conditions of low radiation density the ERE quantum yield is $>>10\%$.
\item The ERE is unpolarized with an isotropic radiation pattern.
\item The ERE peak wavelength and quantum yield are strongly affected by the density 
and hardness of the local UV photon field.
\item The ERE is observed in the absence of scattering and appears therefore not 
associated with sub-micron-sized scattering grains.
\item The ERE appears to be uncorrelated with the strength of the interstellar 2175 
\AA\ extinction feature and the ubiquitous UIR 
emission bands.
\item The ERE carrier particles appear to originate in such dust-forming environments 
as the proto-planetary Red Rectangle nebula and in 
C-rich planetary nebulae.
\end{itemize}
\subsection{Hydrogenated Amorphous Carbon}
Hydrogenated amorphous carbon (HAC) is an organic refractory material consisting of a 
mixture of $sp^2$ and $sp^3$ coordinated 
hydrocarbons with a variable band-gap. The HAC ERE model attracted wide support early 
on, because not only did HAC exhibit PL in apparent 
agreement with observed ERE spectra (Witt \& Schild 1988), it also appears to be an 
interstellar dust component required to explain the 
interstellar 3.4 $\mu$m C-H stretch feature (Pendleton \& Allamandola 2002; Furton, 
Laiho, \& Witt 1999) and is widely employed as an 
integral component of models for interstellar dust. The early HAC PL spectra were 
obtained with exciting radiation near 500 nm
wavelength and they displayed luminescence only in the red region with a spectrum 
similar to that of the ERE. Under UV illumination, 
however, the HAC PL spectrum extends well into the blue and near-UV, consistent with a 
band-gap of over 3 eV (Robertson 1996; Rusli, 
Robertson, \& Amaratunga 1996). Under astrophysical illumination conditions with ample 
UV photons, HAC would therefore  exhibit PL in the 
400 nm to 500 nm spectral range, contrary to observation. The HAC band-gap can be 
narrowed by irradiation or annealing, bringing it 
closer to
the observational constraints, but this also results in a dramatic decrease of the PL 
quantum yield by three or four orders of magnitude. 
As a result, HAC is not able to reproduce the spectral characteristics and the 
observed quantum yield of the ERE simultaneously, and the 
HAC ERE model is no longer considered viable.
\subsection{PAH Molecules and Ions}
Large PAH molecules and their ions are considered likely candidates for explaining the 
ubiquitous UIR emission bands in the ISM and in 
many of the nebular sources in which ERE is observed as well. Given that PAHs are also 
known to fluoresce efficiently (Berlman 1965), 
their role as a potential source of ERE was proposed early on (d'Hendecourt et al. 
1986). Under collision-free conditions and with far-UV 
photons available for excitation, PAHs and their ions are even expected to yield 
several luminescence photons per single excitation, 
resulting in efficiencies of several 100$\%$ (Leger, et al. 1988).
However, neutral PAHs luminesce predominantly in the UV/optical wavelength range (300 
nm - 600 nm) (Birks 1973; Peaden et al. 1980), with 
similar efficiencies in the gas phase (Stockburger 1973; Sgro et al. 2001; Reyle \& 
Brechignac 2000) as in solution (Berlman 1965). 
Studies of the charge state distribution of PAHs in different interstellar 
environments by Bakes \& Tielens (1994) and Weingartner \& 
Draine (2001) agree that a significant fraction of PAHs should still be neutral in 
radiation environments typical of reflection nebulae. 
A size distribution of PAH molecules
would therefore reveal its presence by luminescing not only in the red ERE range but 
in the blue wavelength region as well. The absence 
of detectable PL at wavelengths $<$ 500 nm (Vijh et al. 2003) argues strongly against 
the existence
of smaller PAH molecules (mass $<$ 500 amu) in these environments. Laboratory studies 
of the PL characteristics of larger PAH molecules, 
especially when excited by far-UV photons in a collision-free environment, are needed 
to evaluate the likelihood that PAH molecules are 
the origin of the ERE.

Relatively little is known about the fluorescence characteristics of PAH cations. The 
comparatively small energy difference ($\sim$ 1 eV) 
between the electronic ground state, $\textrm{D}_0$, and the first excited electronic 
state, $\textrm{D}_1$, in most PAH ions (Leach 
1987) causes their potential fluorescence
to occur in the near-IR and not in the optical range (Crawford, Tielens, \& 
Allamandola 1985).
PAH cations are reported to absorb predominantly at energies $>$ 7.75 eV ($<$ 160 nm) 
(Robinson et al. 1997), which results in an 
increased probability of internal conversion to highly excited vibrational levels of 
the electronic ground state D$_0$, from where 
relaxation occurs essentially by emission of IR photons (Leach 1995). This suggests 
that quantum yields for fluorescence by PAH cations 
may be quite low.

The absence of consistent spatial correlations between ERE and UIR band emissions 
attributed to PAHs is not by itself conclusive evidence 
against the PAH origin of ERE, because PAHs in different size ranges, degrees of 
hydrogenation, and charge states, all of which are 
expected to be environment-dependent, could be responsible for UIR bands and ERE to a 
different
extent (Duley 2001). However, so far there is little substantive support \emph{in 
favor} of PAHs as likely candidates for ERE carriers.
\subsection{PAH Clusters}
Spectroscopic data on fluorescence by specific PAH molecules are limited to systems 
with atomic weight $<$ 500 amu or about 40 carbon 
atoms. Duley \& Seahra (1998) suggested that interstellar PAHs exist as much larger 
molecules or
in the form of stacks and aggregates of PAHs of up to 700 carbon atoms,
which could be responsible for the interstellar 2175 \AA\ absorption feature as well 
as the UIR band emissions. They also proposed 
(Seahra \& Duley 1999) that these same PAH clusters will produce PL emission with a 
principal band centered at 700 nm, which would 
account for the ERE. This model suffers from several weaknesses. The model predicts 
two emission side bands, with peaks at 500 nm and 
1000 nm wavelength, to be associated with the main ERE band.
A search for these additional bands in two of the strongest ERE filaments in NGC 7023 
by Gordon et al. (2000) failed to produce evidence 
for either one.
The model predicts a remarkably constant peak wavelength for the main ERE band
for a wide range of size distributions of PAH clusters, contrary to the observational 
evidence (Figure 6). Finally, the model predicts 
correlations between the strengths of the 2175 \AA\ band and the ERE band, which is 
not supported by observations. The most attractive 
feature of this model is the efficient utilization of the same dust component to 
explain several interstellar
phenomena, namely the 2175 \AA\ band, ERE, UIR bands. It would be highly desirable
to obtain laboratory data on PAH clusters which could support the multiple claims made 
for this model (Duley 2001).
\subsection{Other Carbonaceous Carrier Models}
The $\textrm{C}_{60}$ molecule exhibits a PL spectrum resembling that of the ERE in a 
few reflection nebulae, leading Webster (1993) to 
suggest $\textrm{C}_{60}$
as the source of ERE. However, sensitive searches for the presence of 
$\textrm{C}_{60}$
and $\textrm{C}_{60}^+$ in interstellar space and in NGC 7023 (Snow \& Seab 1989; 
Moutou et al. 1999; Herbig 2000) failed to find 
observable traces of this species. Furthermore, the measured quantum yield of 
$\textrm{C}_{60}$ PL is only 8.5$\times$$10^{-4}$ (Kim et 
al. 1992), missing the ERE requirement by several orders of magnitude.

Discharges through gas mixtures containing hydrocarbons result in non-volatile 
carbonaceous residues with interesting PL characteristics. 
Wdowiak et al. (1989), using a gas mixture containing CO, $\textrm{CH}_4$, 
$\textrm{N}_2$, $\textrm{H}_2\textrm{O}$, and Ar, with 
relative abundances approximating those believed to apply to the Red Rectangle, 
produced luminescing residues of high quantum yield. 
However, under UV illumination the PL spectrum consistently peaked at wavelengths $<$ 
600 nm, with substantial parts of the luminescence 
spectrum extending through the 400 - 500 nm range. The spectral mismatch with the 
observations plus the presence of blue PL appear to 
rule out this particular candidate.

Quenched carbonaceous composite (QCC), a filmy residue produced through a microwave 
discharge in a low-density methane plasma by Sakata 
et al. (1992), was shown to be a more promising ERE carrier candidate. This material
exhibits PL spectra with variable peak wavelengths spanning the 680 - 725 nm range, 
depending on the substrate temperature during 
deposition. 
Blue luminescence is absent in the presence of UV illumination, in agreement with 
observations. Mass spectroscopy of QCC (Sakata et al. 
1983) indicates that QCC does not contain large PAH molecules but appears to consist 
of a random agglomeration of 1- to 4-ring PAHs, 
small aliphatic molecules and radicals.
When dissolved in liquid freon, QCC exhibits blue luminescence, reminiscent of
the fluorescence of small PAHs. A more complete characterization of QCC through 
additional laboratory work would clearly be desirable.
\subsection{Silicon Nanoparticles}
In recent years, silicon nanoparticles (SNPs) (Witt et al. 1998; Ledoux et al. 1998, 
2000, 2001, 2002) have emerged as an interesting 
alternative to the carbonaceous ERE carrier candidates discussed above. The SNP model 
meets the observational constraints posed by ERE observations 
in respect to spectral variability and quantum yield better than most of the 
carbonaceous candidates. The photophysics of SNPs (Smith \& 
Witt 2002) also appears to be consistent with the observed variations of ERE peak 
wavelength and ERE quantum yield with environmental 
conditions. The balance between photo-ionization and recombination with free 
electrons, combined with the fact that charged SNPs are much 
less likely to luminesce, explains the variation of the ERE quantum yield with 
UV-radiation density, while the expected photo 
fragmentation of multiply charged SNPs leads to the erosion of the SNP size 
distribution, starting with the smallest sizes, and a 
corresponding shift to much longer ERE peak wavelengths in astrophysical environments 
with the highest radiation densities. Studies of 
the optical characteristics of SNPs (Amans et al. 2003) show the absorption 
coefficient still rising at the wavelength of 200 nm, where 
current measurements stop.
This is consistent with ERE excitation requirements, but measurements at shorter 
wavelengths are urgently needed.

The highly efficient PL exhibited by SNPs is the results of quantum confinement in 
nanocrystals and passivation of surface dangling bonds 
with atoms such as H, O, N, C, and Fe. Most existing laboratory studies have been done 
with O- and H-passivation, and it has been 
found that the surface composition affects the PL spectrum (Wolkin et al. 1999; Zhou, 
Brus, \& Friesner 2003), in the sense that SNPs 
with H-passivation and diameters of $<$ 2.5 nm luminesce at blue and near-UV 
wavelengths, while O-passivated SNPs luminesce in the red 
only. This argues against H-passivated SNPs as ERE carriers, because blue PL is not 
observed
under interstellar conditions. Also not observed is an interstellar absorption feature 
at 4.6 $\mu$m, which would have resulting from the 
expected Si-H vibrational transition. For these reasons, Witt et al. (1998) suggested 
that interstellar SNPs would most likely be 
O-passivated. Once produced in a circumstellar outflow, SNPs with oxygen passivation 
would persist even in the ISM with abundant H-atoms, 
because the Si-O bond is energetically two to three times stronger than the Si-H bond. 
Such particles would contribute to the widely 
observed 9.7 $\mu$m vibrational Si-O absorption band, generally referred to as the 
``silicate'' feature. However, Li \& Draine (2002) have pointed 
out that stochastically heated O-passivated SNPs would produce a 20 $\mu$m emission 
band well in excess of currently established 
observational limits. This objection might be overcome, if SNPs with O-passivation 
were either attached to or embedded in larger grains.
Experiments, in which SNPs were produced by ion implantation in solids with subsequent 
annealing, have demonstrated that SNPs preserve 
their luminescent characteristics even when embedded (e.g. Iwayama et al. 2002). 
Another way to
overcome the objection by Li \& Draine (2002) is to consider other forms of 
passivation. For example, recent laboratory work (Mavi et al. 
2003) has shown that passivation with Fe-atoms yields even more persistent and 
efficient photoluminescence in SNPs than O-passivation.
Among astrophysical environments in which Fe-passivated SNPs might form are S-stars 
(Ferrarotti \& Gail 2002), which are characterized by 
an atmospheric abundance ratio C/O $\approx$ 1. That such environments are likely to 
contribute grains consisting of Si and Fe to the 
interstellar dust mixture is supported by the recent observation of a new 1.5 $\mu$m 
dust emission feature, attributed to 
$\beta$-$\textrm{FeSi}_2$ particles (Gordon et al. 2000). Carbon should be able to 
provide passivation equally well, although experiments in this direction have yet to 
be done. The existence of Fe- or C-passivated SNPs could also overcome the 
observational 
constraint, which so far
has restricted ERE detections to C-rich environments among planetary nebulae
(Furton et al. 1992).

A critical observational test of the SNP hypothesis consists of successful searches 
for a SNP defect band observable near 1.15 $\mu$m 
(Gordon et al. 2000). If additional detections of this near-IR band are made, one 
needs to test the correlation between the peak 
wavelength of the main SNP band and that of the near-IR band (Hill \& Whaley 1996), 
which is well-established through laboratory 
experiments. An additional test is possible in the form of searches for enhanced 
emission from gas-phase [SiII] at 34.8 $\mu$m in regions 
where
SNPs are expected to be destroyed by photo-fragmentation, e.g. between 
photo-dissociation fronts and their illuminating sources (Fuente 
et al. 2000).
The strong enhancement of such [SiII] emission in the central part on NGC 7023,
a region where ERE is essentially absent, is consistent with photo-destruction
of SNPs in this region. Photo-destruction of silicates would also lead to an increase 
in gas-phase SiII, which, however, should then be 
associated with corresponding increases in the abundances of gas-phase Fe and Mg.

In summary, the SNP model, while promising, also suffers from a number of loose ends.
The study of the optical properties of SNPs must be pushed into the vacuum 
ultraviolet, the full range of possible passivation schemes must be explored, the 
process of SNP formation in stellar outflows must be examined, and the question of the 
long-term persistence of the PL of SNPs under astrophysical conditions over 
astronomical times scales must be investigated.
\section{Conclusions}
The ERE is a well-characterized property of dust existing in a wide range of 
astrophysical environments. The carrier of the ERE is a 
major component of interstellar dust: it intercepts $\sim$20$\%$ of the UV/optical 
photons in the diffuse ISM, if the intrinsic quantum 
yield is $\sim$50$\%$, and a still higher fraction of the UV/optical photons, if the 
yield is lower than $\sim$50$\%$.. Yet, no current 
comprehensive model for interstellar dust explicitly accounts for the existence of 
the ERE. The identification of the ERE carrier 
remains a challenge for laboratory astrophysics, dust theory, and astronomical 
observation.
\begin{acknowledgements}
Past and current collaborators have contributed substantially to the our present 
understanding of this subject: Todd Boroson, Martin 
Cohen, Doug Furton, Karl Gordon, Daniele Pierini, Dmitri Ryutov, Jr., Rudy Schild, and 
Tracy Smith.
We acknowledge their contributions with gratitude. We also thank Jean-Pierre Sivan for 
providing Figure 5 and Astronomy \& Astrophysysics 
for permission to reproduce it. Finally, we acknowledge NASA Grant NAG5-9262 for 
financial support.
\end{acknowledgements}

\end{document}